# Non-Point Visible Light Transmitter Localization based on Monocular Camera

Hongxiu Zhao[1], Xun Zhang[1], Faouzi Bader[1], Yue Zhang[2]
[1]ECoS-LISITE, Institut Supérieur d'Électronique de Paris, [2]University of Leicester
[1]Paris, France, [2]Leicester, UK
hongxiu.zhao@ext.isep.fr

***Abstract*—Many algorithms for visible light positioning (VLP) localization do not consider the shapes of the transmitters, which leads to the impracticality of the algorithm and the low localization accuracy. Therefore, this paper proposes a novel VLP algorithm and addresses the problems in terms of practicality and complexity by using one non-point transmitter based on a monocular. Because the shape of the transmitter is considered, the proposed algorithm is easy for practice and has wide applicability. Besides, it decreases the computation for simple geometric model and expands the coverage for there is a greater chance of receiving signals from one light than that of receiving signals from multiple lights.**

## I. Introduction

As German proposed in 2013, the 4.0 industrial revolution would influence the world with massive devices and actuators in multiple applications and scenarios. According to Cisco forecasting, by 2030, the predicted number of connected devices on the Internet will reach up to 500 million. Under this situation, coupled with the effect of Beyond 5 Generation (B5G) and the expected features of 6 Generation (6G), there is a growing demand for massive Device-to-Device (D2D) communication on the Internet of Things (IoT), thereby, the standards of the communication performances are getting stricter and they may depend on the performance of localization under certain circumstances, especially indoor localization. There already exist many indoor localization technologies including WLAN, Bluetooth, UWB, VLP, etc. Considering the coverage, cost of infrastructure, localization precision and some non-ignorable factors such as information security and abundant spectrum resources, visible light positioning (VLP) is adopted to accomplish better indoor localization. Among the existing localization methods including Time of Arrival (TOA), Difference Time of Arrival (DTOA), Angle of Arrival (AOA), Received Signal Strength (RSS), camera and fingerprint method, camera localization is a lower cost and higher speed method with the simper operation[1]. Although existing monocular image-based 3D object detection methods have many problems to accomplish 3D localization, they are favored because of the simplicity of set up and suitability for inexpensive robots detecting and moving. Compared with binocular camera, monocular camera is highly appreciated due to the low cost, high precision, considerable flexibility, low dependency on the environment and equipment, and the simple calibration[2-4]. Therefore, in this article, monocular camera is utilized to accomplish the efficient localization. Monocular localization algorithms are categorized into algorithm based on single frame image, the one based on double frame, and the one based on multiple image. In reality, the last two algorithm are complicated and hard to achieve high accuracy 3D localization, therefore, many researches just focus on the algorithm based on single frame image. However, this algorithm has to handle the massive data and the corresponding updates with the projection of the featured points, straight line, and curve line, on the image plane, which leads to still complex computation[5-6]. Therefore, a novel algorithm based on monocular camera is proposed in this paper, which could deal with the problems mentioned above.

Meanwhile, although there are many algorithms for VLP localization, almost none of them considers the shapes of the transmitters[7-9], which leads to the impracticality of the algorithm and the low localization accuracy. Most researches consider those shaped transmitters as the point-transmitters, limiting the application of localization. Therefore, in this paper, a non-point transmitter will be considered and utilized. In summary, this paper proposes a novel VLP algorithm and addresses the problems in terms of practicality and complexity by using one transmitter based on a monocular.

The rest of paper will be organized as follows. Section II describes the system model and the details are demonstrated in Section III. Simulation results are presented in Section IV. Finally, Section V concludes the paper.

## II. System model

In reality, there are many different shapes of flat lights, such as round light and rectangular light. Unlike the existing researches only considering the point-transmitter, the proposed algorithm is suitable for localization based on lights with various shapes. Supposing that there are lots of transmitters (lights) on the ceiling of the room and each one has their own identification (ID), the camera could recognize which transmitter is the one for its photo according to the ID information of each transmitter. Although there are many kinds of shaped transmitters and the proposed algorithm is suitable for almost all, this model takes the rectangular one as an example for the clear elaboration. The receiver camera is equipped on the mobile. The overall model and the world coordinate are as shown in figure 1(a). H means the height of the room whose ceiling is equipped with rectangular transmitters.

In the proposed algorithm, the non-point transmitters are considered as a collection of many evenly distributed virtual points. Therefore, in this Section, the rectangular transmitter is considered as a collection of massive uniformly distributed points as shown in figure 1(b). It is assumed that the localization of the center point of the transmitter and the size of light is known and the monocular camera could receive the ID information of the transmitter, recognize the points of it and pick up some featured points among them, such as point A, B, and C, as shown in Figure 1 (b). Besides, the required camera calibration has been accomplished.

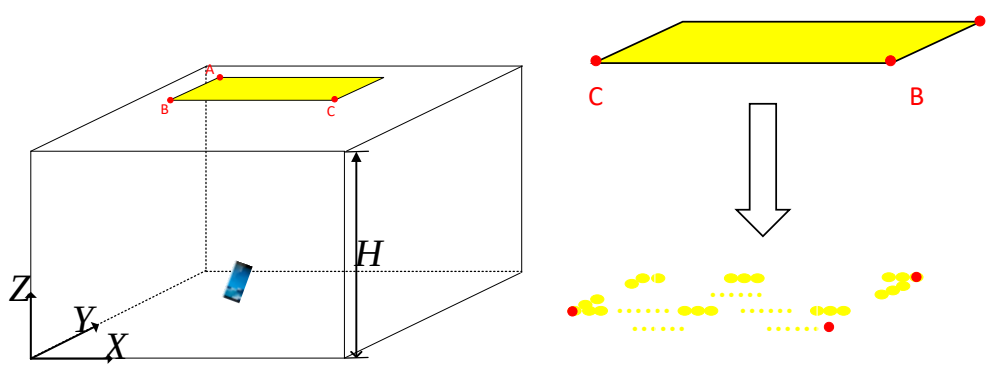

(a) (b)

Fig. 1. (a) System model. (b) Equivalent model of the transmitter.

## III. PROPOSED ALGORITHM

### A. Foundation

Considering the shapes of the transmitters, this paper would propose a novel algorithm just based on one monocular and one non-point transmitter. This algorithm

Firstly, the imaging principle can be demonstrated[10-11]. As shown in Figure 2, there are four coordinate systems, including world coordinate with $O_w$, camera coordinate with $O_c$, image coordinate with O, and pixel coordinate with $O_{uv}$. The world coordinate could be transformed into the camera coordinate when there is orientations and displacement.

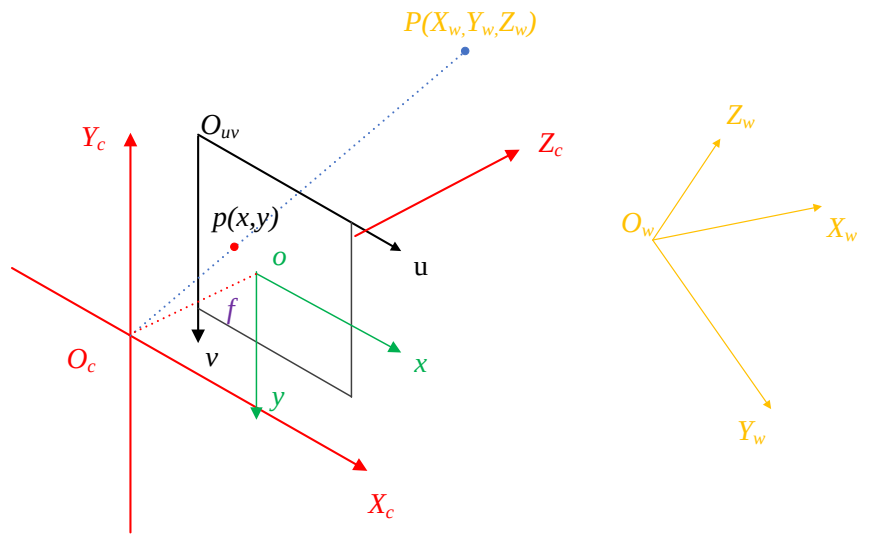

Fig. 2. Imaging priciple

When the world coordinate rotates $\theta$ around the z axis, there will be the rotation matrix $R_z$ depicted as (1). Similarly, the rotation around x and y axis, $R_x$ and $R_y$ could be obtained in (2) and (3), respectively.

$$R_z = \begin{pmatrix} \cos(\theta) & -\sin(\theta) & 0 \\ \sin(\theta) & \cos(\theta) & 0 \\ 0 & 0 & 1 \end{pmatrix} \tag{1}$$

$$R_x = \begin{pmatrix} 1 & 0 & 0 \\ 0 & \cos(\varphi) & \sin(\varphi) \\ 0 & -\sin(\varphi) & \cos(\varphi) \end{pmatrix} \tag{2}$$

$$R_y = \begin{pmatrix} \cos(w) & 0 & -\sin(w) \\ 0 & 1 & 0 \\ \sin(w) & 0 & \cos(w) \end{pmatrix} \tag{3}$$

Therefore, the reference point ($x_w$, $y_w$, $z_w$) in world coordinate is transformed into the point ($x_c$, $y_c$, $z_c$) in camera coordinate as described in (4).

$$\begin{pmatrix} x_c \\ y_c \\ z_c \end{pmatrix} = R \cdot \begin{pmatrix} x_w \\ y_w \\ z_w \end{pmatrix} \tag{4}$$

Where $R = R_z \cdot R_x \cdot R_y$.When there is a displacement T between two coordinates, (5) could describe the relationship.

$$\begin{pmatrix} x_c \\ y_c \\ z_c \\ 1 \end{pmatrix} = \begin{pmatrix} R & T \\ 0 & 1 \end{pmatrix} \cdot \begin{pmatrix} x_w \\ y_w \\ z_w \\ 1 \end{pmatrix} \tag{5}$$

Where T is a three-dimensional vector, implicating the distance between the origins of two coordinates.

The transformation between the image coordinate ($x_i$, $y_i$) and the pixel coordinate (u, v) is demonstrated in (6).

$$\begin{pmatrix} u \\ v \\ 1 \end{pmatrix} = \begin{pmatrix} 1/d_x & 0 & u_0 \\ 0 & 1/d_y & v_0 \\ 0 & 0 & 1 \end{pmatrix} \cdot \begin{pmatrix} x_i \\ y_i \\ 1 \end{pmatrix} \tag{6}$$

Where ($u_0$, $v_0$) is the origin of the image coordinate in the pixel coordinate system and ($d_x$, $d_y$) is the pixel size along x and y axis. Therefore, the overall transformation between the world coordinate and the pixel coordinate could be clarified by (7).

$$\begin{pmatrix} u \\ v \\ 1 \end{pmatrix} = \begin{pmatrix} 1/d_x & 0 & u_0 \\ 0 & 1/d_y & v_0 \\ 0 & 0 & 1 \end{pmatrix} \cdot \begin{pmatrix} f & 0 & 0 & 0 \\ 0 & f & 0 & 0 \\ 0 & 0 & 1 & 0 \end{pmatrix} \cdot \begin{pmatrix} R & T \\ 0 & 1 \end{pmatrix} \cdot \begin{pmatrix} x_w \\ y_w \\ z_w \\ 1 \end{pmatrix} \tag{7}$$

Where f is the focal length of the camera. Usually, (7) could be simplified as (8).

$$z_c \begin{pmatrix} u \\ v \\ 1 \end{pmatrix} = \begin{pmatrix} f_x & 0 & u_0 & 0 \\ 0 & f_y & v_0 & 0 \\ 0 & 0 & 1 & 0 \end{pmatrix} \cdot \begin{pmatrix} R & T \\ 0 & 1 \end{pmatrix} \cdot \begin{pmatrix} x_w \\ y_w \\ z_w \\ 1 \end{pmatrix} \tag{8}$$

According to what mentioned above, the relationship between each distance could be found and utilized. Therefore, the proposed algorithm is based on the geometric transformation. The details are demonstrated as follows.

### B. Proposed algorithm

It is assumed that the localization of the center *E* point of the transmitter and the size of the light has already been known, therefore, the localization of every virtual point, such as points on the edge of the light, *A*, *B*, *C*, and *D*, can be calculated and set in the database. After the snapshot, the camera could know the localizations of all needed points including *A*, *B*, *C*, and *D*. Meanwhile, their mapping on the image plane is also achieved, creating points *a*, *b*, *c*, and *d*, shown as figure 2(a), where *P* is the optical center of the camera and considered as the localization of the mobile phone.

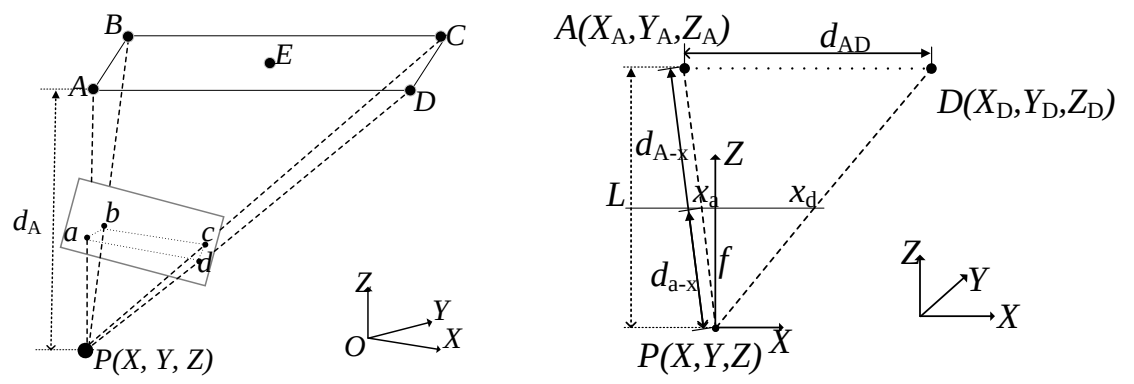

(a) (b)

Fig. 3. (a) Mapping of the transmitter. (b) Front view of the imaging model.

To accomplish the localization of the camera, trilateral geometry localization is applied. The distances between the optical center point and several identified points of the light are supposed to be calculated, such as $d_A$ shown in Figure 2(a). The process for the distance $d_A$ is described below.

Firstly, according to the triangle theorem, (1) could be obtained.

$$\frac{f}{L} = \frac{|x_d - x_a|}{d_{AD}} \tag{1}$$

Where $L$ is the vertical distance between the camera and transmitter plane, $f$ is the focal distance, $|x_d - x_a|$ that could be calculated after the camera calibration is the distance between two image points and $d_{AD}$ can be obtained in advance. Therefore, $L$ can be worked out as (2).

$$L = \frac{d_{AD}}{|x_d - x_a|} \cdot f \tag{2}$$

At the same time, (3) and (4) also works.

$${d_{a-x}}^2 = f^2 + |x_a|^2 \tag{3}$$

$${d_a}^2 = {d_{a-x}}^2 + |y_a|^2 \tag{4}$$

Where $|x_a|$ is the projection distance of the one between the image point a and the center point of the image plane on the x-axis, and $|y_a|$ is the projection distance on y-axis. $d_{a-x}$ is shown in Figure 2(b). $d_a$ refers to the distance between the optical center of the camera and the mapping point a on the image plane. It can be noticed that Figure 2(b) is the front view which can only indicate the geometric relationship of YOZ section plane while $d_a$ is supposed to be the distance existing in three-dimensional space, the reason for (4).

Based on $d_a$, the According to (5), the distance $d_A$ is calculated.

$$\frac{f}{L} = \frac{d_a}{d_A} \tag{5}$$

Similarly, $d_B$ and $d_C$ could be obtained. By localization method such as least square method, the final localization of the camera is accomplished. In this paper, the trilateral geometry localization is applied for the less computation by (5) – (7).

$$d_A{}^2 = (X - x_A)^2 + (Y - y_A)^2 + (Z - z_A)^2 \tag{5}$$

$$d_B{}^2 = (X - x_B)^2 + (Y - y_B)^2 + (Z - z_B)^2 \tag{6}$$

$$d_C{}^2 = (X - x_C)^2 + (Y - y_C)^2 + (Z - z_C)^2 \tag{7}$$

Where $(x_A, y_A, z_A)$, $(x_B, y_B, z_B)$, and $(x_C, y_C, z_C)$ are the positions of the needed points A, B, and C of the light. shown in figure 2(a). $z_A = z_B = z_C = H$ and $Z = H - L$.

According to the algorithm above, $(X, Y, Z)$ is finally worked out.

## IV. SIMULATION RESULT

The proposed algorithm could be verified by the following simulations. Firstly, simulating an environment as shown in Fig. 4. The width of the simulated room is 3m, the length of it is also 3m, and the height is the 5m. As described in chapter II, the algorithm is practical and fit for non-point light, therefore, a rectangular is taken as an example in this part. It is fixed on the ceiling of the room and the coordinates of its four corners are (1,1,5), (1,2,5), (2,1,5), (2,2,5), respectively. The known points are placed on in z=2 plane with different distances from 0 to 3 m by 0 to 3 m with 500 cm increasing steps, as shown in Fig.1.

Based on a flexible new technique for camera calibration, camera parameters could be obtained and depicted in table 1. The parameter intrinsic matrix in the table contains the focal length and origin of the image coordinate in the pixel coordinate system, as described in (8).

TABLE 1 camera parameters

| | |
|---|---|
| Focal length (um) | $[4.0001 \cdot 10^3, 4.0102 \cdot 10^3]$ |
| Principle point (um) | $[2.6348 \cdot 10^3, 1.5286 \cdot 10^3]$ |
| Intrinsic matrix | $\begin{bmatrix} 4.0001 \cdot 10^3 & 0 & 0 \\ 0 & 4.0102 \cdot 10^3 & 0 \\ 2.6348 \cdot 10^3 & 1.5286 \cdot 10^3 & 1 \end{bmatrix}$ |

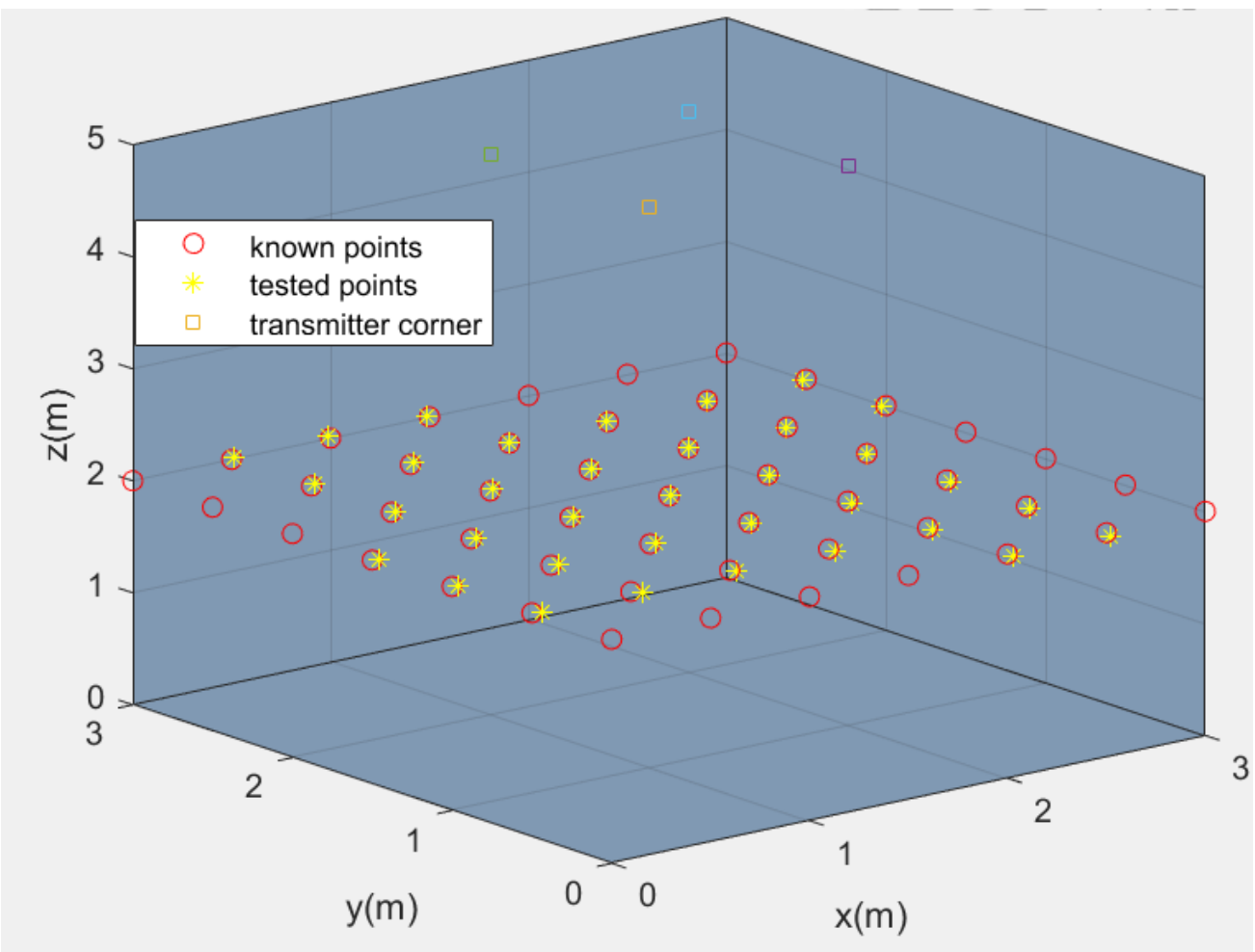

Fig. 4. Known points and calculated points in the simulated room

Fig.1 proves the feasibility of the proposed algorithm because the positions of the calculated points are close to ones of known points. To exhibit the subtle differences of them, we will show the calculated points from the perspective of XoY plane and YoZ plane, respectively. The view from the XoY plane verifies the accuracy of the two-dimensional positioning, Fig. 5. shows that the heights of the known points are 2m and the calculated points are close to them and the maximum offset is 0.05m.

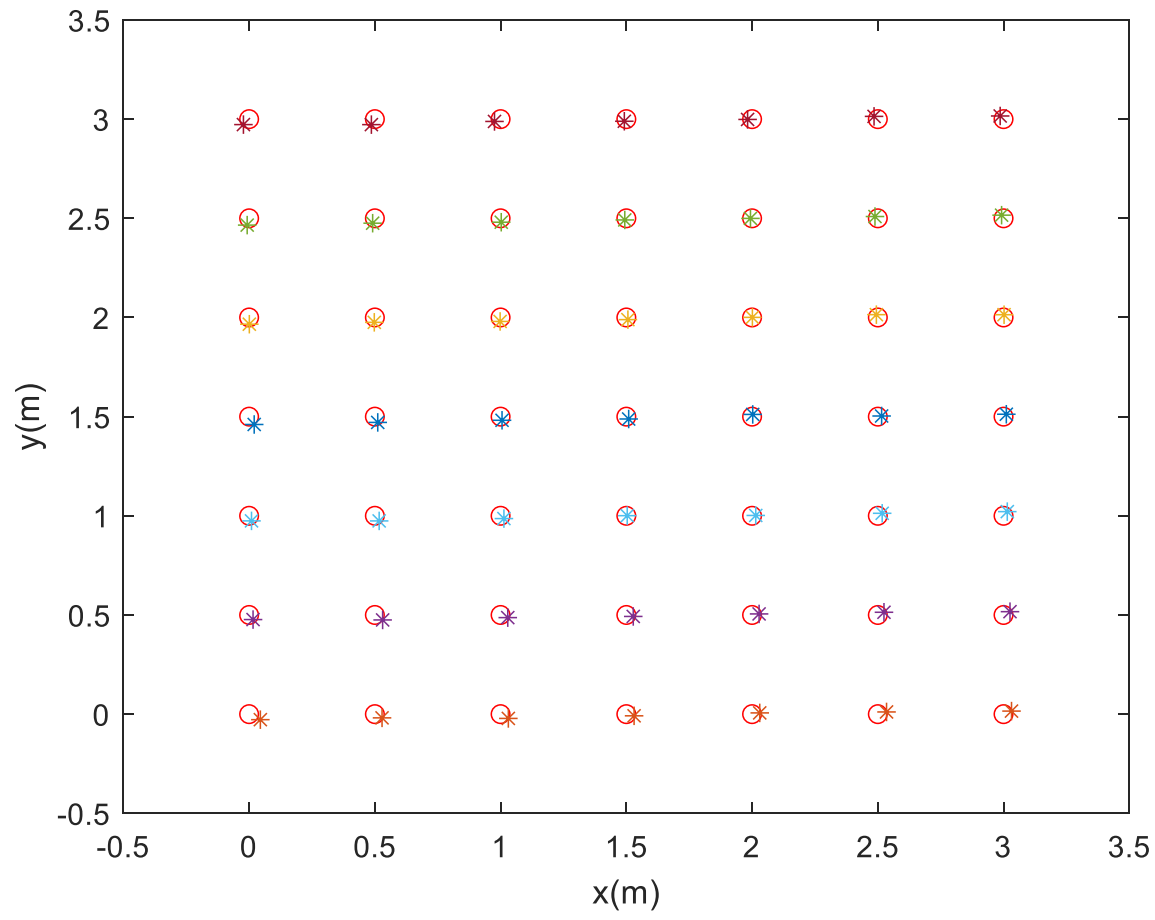

Fig. 5. Known points and calculated points on XoY plane

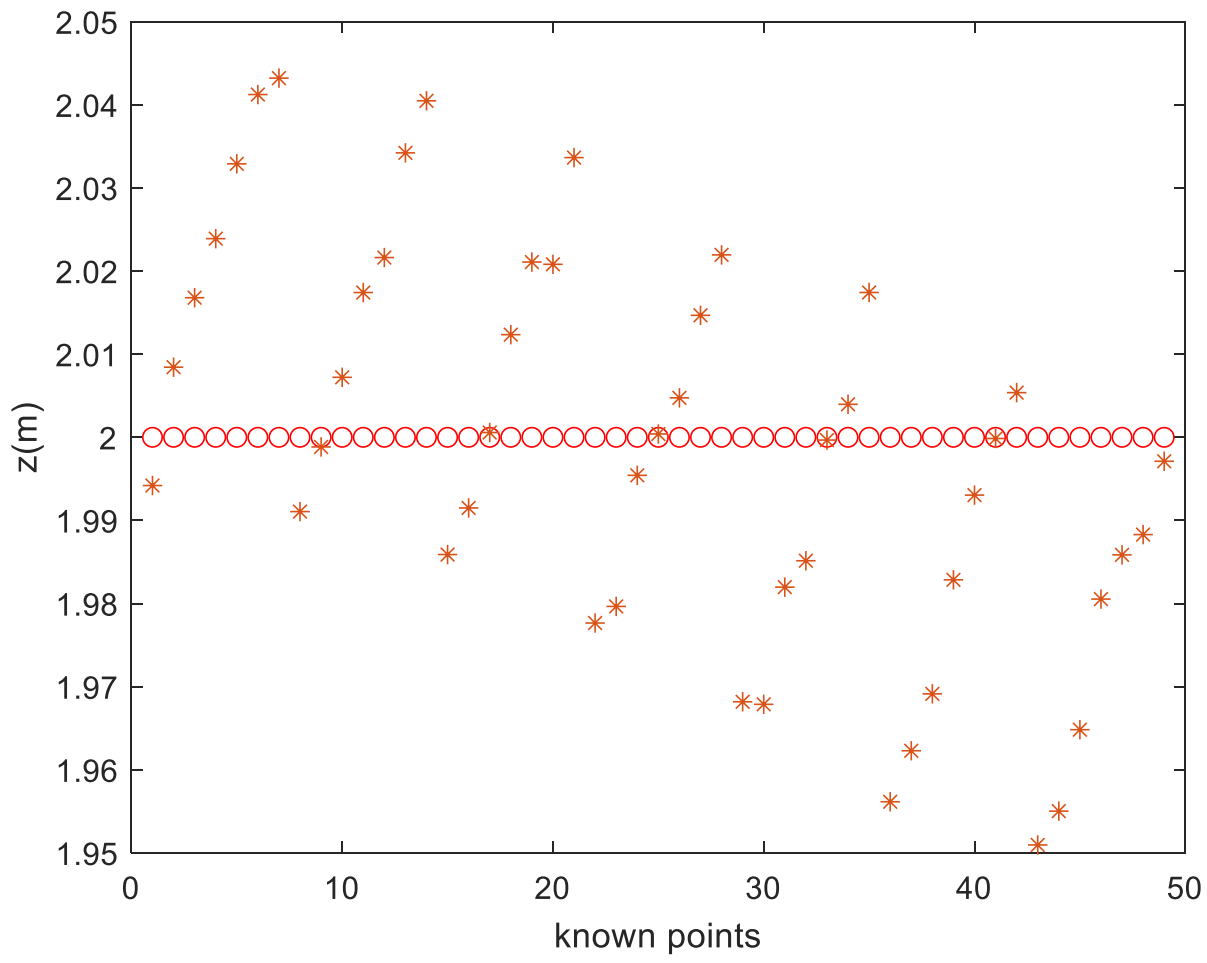

Fig. 6. Known points and calculated points on YoZ plane

Fig. 7 shows the cumulative distribution function (CDF) of the root mean squared error (RMSE) of the calculated points on the XoY plane and the YoZ plane. This verifies the convergence of the algorithm that can achieve the centimeter accuracy. The maximum RMSE is 0.06m. Besides from the perspective of XoY plane and YoZ plane, RMSE could converge faster.

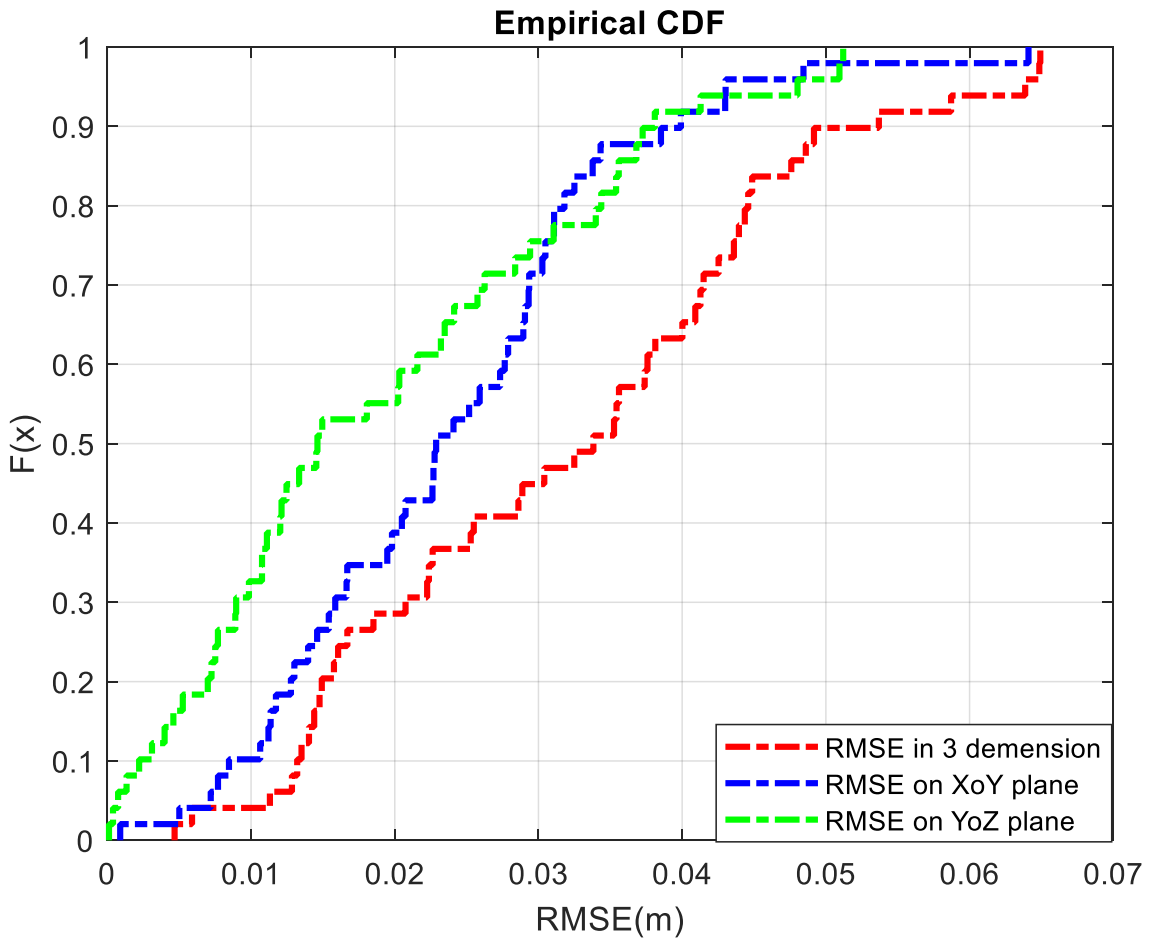

Fig. 7. Known points and calculated points on YoZ plane

## V. Conclusion

In this paper, the proposed algorithm is only based on one transmitter by considering the shapes of the light with monocular camera. The necessary data in advance are the position of the middle point $E$ of the transmitter and the size of it. The camera takes a snapshot and recognize the light's ID, then, according to the known data, geometric image model can be constructed for the transmitter, image plane and optical center of the camera. After the calculation of distances between the optical center of the camera and the needed points of the transmitter, trilateral positioning method is exhibited for the final localization. Because the shape of the transmitter is considered, the proposed algorithm is easy for practice and has wide applicability. Besides, it decreases the computation for simple geometric model and expands the coverage for there is a greater chance of receiving signals from one light than that of receiving signals from multiple lights.

## Acknowledgement

The authors gratefully acknowledge the financial supports of the Chinese Scholarship Council and the EU Horizon 2020 program towards the 6G BRAINS project H2020-ICT 101017226.